\documentclass[sigconf]{acmart}
\usepackage{hyperref}
\pdfoutput=1
\renewcommand\footnotetextcopyrightpermission[1]{} 
\newtheorem{lemma}{Lemma}
\usepackage{times,amsmath,epsfig}
\usepackage{epstopdf}
\usepackage{caption}
\usepackage{algorithm}
\usepackage{algpseudocode}
\usepackage{amsmath}
\usepackage{graphicx}
\usepackage{url}
\usepackage{multirow}
\usepackage{multicol}
\usepackage{subfigure}
\usepackage{threeparttable}
\usepackage{booktabs}
\usepackage{amsfonts}
\usepackage[framemethod=1]{mdframed}
\usepackage{enumitem}

\definecolor{shadecolor}{rgb}{0.878906, 0.878906, 0.878906}

\ifodd1
    
    \newcommand{\com}[1]{\textbf{\color{blue} (COMMENT: #1)}} 
\else
    
    \newcommand{\com}[1]{}
\fi

\newcommand{\tabincell}[2]{\begin{tabular}{@{}#1@{}}#2\end{tabular}}
\def\BibTeX{{\rm B\kern-.05em{\sc i\kern-.025em b}\kern-.08emT\kern-.1667em\lower.7ex\hbox{E}\kern-.125emX}}
    
\keywords{Deep learning, Deep neural network, Optimal power flow.}

\title{DeepOPF: Deep Neural Network for DC Optimal Power Flow}
\author{Xiang Pan}
\affiliation{%
 \department{Information Engineering}
 \institution{The Chinese University of Hong Kong}
}

\author{Tianyu Zhao}
\affiliation{%
 \department{Information Engineering}
 \institution{The Chinese University of Hong Kong}
}

\author{Minghua Chen}
\affiliation{%
 \department{Information Engineering}
 \institution{The Chinese University of Hong Kong}
}

\begin{abstract}
We develop \textsf{DeepOPF} as a Deep Neural Network (DNN) approach for solving direct current optimal power flow (DC-OPF) problems. 
\textsf{DeepOPF} is inspired by the observation that solving DC-OPF for a given power network is equivalent to characterizing a high-dimensional mapping between the load inputs and the dispatch and transmission decisions. We construct and train a DNN model to learn such mapping, then we apply it to obtain optimized operating decisions upon arbitrary load inputs. We adopt uniform sampling to address the over-fitting problem common in generic DNN approaches. We leverage on a useful structure in DC-OPF to significantly reduce the mapping dimension, subsequently cutting down the size of our DNN model and the amount of training data/time needed. We also design a post-processing procedure to ensure the feasibility of the obtained solution. Simulation results of IEEE test cases show that \textsf{DeepOPF} always generates feasible solutions with negligible optimality loss, while speeding up the computing time by \textit{two orders of magnitude} as compared to conventional approaches implemented in a state-of-the-art solver. 
\end{abstract}

\begin{document}

\maketitle

\section{Introduction}\label{sec:intro}
The ``deep learning revolution'' largely enlightened by the October 2012 ImageNet victory~\cite{krizhevsky2012imagenet} has transformed various industries in human society, including artificial intelligence, health care, online advertising, transportation, and robotics. As the most widely-used and mature model in deep learning, Deep Neural Network (DNN) \cite{goodfellow2016deepma}  demonstrates superb performance in complex engineering tasks such as recommendation \cite{covington2016deep}, bio-informatics \cite{bty543}, mastering difficult game like Go \cite{silver2016mastering}, and human pose estimation \cite{6909610}. 
The capability of approximating continuous mappings and the desirable scalability make DNN a favorable choice in the arsenal of solving large-scale optimization and decision problems in various engineering systems.  In this paper, we apply deep learning to power systems and develop a DNN approach for solving the essential optimal power flow (OPF) problem in power system operation.



The OPF problem, first posed by Carpentier in 1962 in~\cite{carpentier1962contribution}, is to minimize an objective function, such as the cost of power generation, subject to all physical, operational, and technical constraints, by optimizing the dispatch and transmission decisions. These constraints include Kirchhoff's laws, operating limits of generators, voltage levels, and loading limits of transmission lines~\cite{johnson1989electric}. The OPF problem is central to power system operations as it underpins various applications including economic dispatch, unit commitment, stability and reliability assessment, and demand response. While OPF with a full AC power flow formulation (AC-OPF) is most accurate, it is a non-convex problem and its complexity obscure practicability. Meanwhile, based on linearized power flows, DC-OPF is a convex problem admitting a wide variety of applications, including electricity market clearing and power transmission management. See e.g.,~\cite{frank2012optimal1, frank2012optimal2} for a survey.

Our DNN approach is inspired by the following observations on the  characteristics of OPF and its application in practice. 
\begin{itemize}
	\item Given a power network, solving the OPT problem is equivalent to depicting a high-dimensional mapping between load inputs and optimized dispatch and transmission decisions.
	
	\item In practice, the OPF problem is usually solved repeatedly for the same power network, e.g., every five minutes by CAISO, with different load inputs at different time epochs. 
\end{itemize}
As such, it is conceivable to leverage the universal approximation capability of deep feed-forward neural networks~\cite{hornik1991approximation,karg2018efficient}, to learn such mapping for a given power network, and then apply the mapping to obtain operating decisions upon giving load inputs (e.g., once every five minutes). 


We develop \textsf{DeepOPF} as a DNN approach for solving DC-OPF problems. As compared to conventional approaches based on interior-point methods~\cite{780938}, \textsf{DeepOPF} excels in (i) reducing computing time and (ii) scaling well with the problem size. These salient features are particularly appealing for solving the (large-scale) security-constrained DC-OPF problem, which is central to secure power system operation with contingency in consideration. Note that the complexity of constructing and training a DNN model is minor if amortized over the many DC-OPF instances (e.g., one per every five minutes) that can be solved using the same model.  Our main contributions are summarized in the following.

First, after reviewing the OPF problem in Sec.~\ref{sec:OPF.review}, we propose a DNN framework for solving DC-OPF problem, \textsf{DeepOPF}, in Sec.~\ref{sec:DeepOPF}. \textsf{DeepOPF} integrates the structure of the DC-OPF problem into the design of the DNN model and the loss function to be used in the learning process. The framework also includes a pre-processing procedure to calibrate the inputs to improve DNN training efficiency and a post-processing procedure to ensure the feasibility of the solutions obtained from the DNN model. 

Second, as described in Sec.~\ref{ssec:load.sampling.and.pre-preprocessing}, we adopt uniform sampling to address the over-fitting problem common in generic DNN approaches. Furthermore, as discussed in Sec.~\ref{ssec:linear.transformation.and.dimension.reduction}, we  leverage on a unique structure in DC-OPF to significantly reduce the mapping dimension, subsequently cutting down the size of our DNN model and the amount of training data/time needed. 

Finally, we carry out simulation using pypower \cite{tpcwTrey1} and summarize the results in Sec.~\ref{sec:simulations}. Simulation results of IEEE test cases show that \textsf{DeepOPF} always generates feasible solutions with negligible optimality loss, while speeding up the computing time by \textit{two orders of magnitude} as compared to conventional approaches. The DeepOPF approach is applicable to more general settings, including the large-scale security-constrained OPF and non-convex AC-OPF, which we leave for future studies. 

\subsection{Related Work}
Generally, the OPF problems can be divided into three forms: economic dispatch (ED), DC-OPF and AC-OPF problems \cite{823997}. The AC-OPF problem is the original OPF problem, which is non-convex and usually difficult to handle. Thus in practical applications, some work focus on solving the relaxed version of the AC-OPF problem like ED and DC-OPF problems since they are easier to solve and the solutions of these problems are useful in the analysis of the AC-OPF problem. ED and DC-OPF problems \cite{4956966} are obtained by removing or linearizing some constraints in the AC-OPF problem, respectively.

The methods for solving the OPF problem are mainly divided into three categories. One is the methods based on numerical iteration algorithms. The OPF problem to be solved can be first approximated as an optimization problem like semi-definite programming \cite{Bai2008Semidefinite}, quadratic programming \cite{71294}, second-order cone programming \cite{1664986} or linear programming  \cite{6756976}, and the numerical iteration solvers like interior-point methods \cite{4538518}, \cite{5575435} were applied to obtain the optimal solutions. However, the time complexity of these numerical-iteration based algorithms may be substantial. Usually the computation time increases when the scale of transmission power system becomes large. Compared with these methods, the proposed DNN-based approach can reduce the processing time during calculation as the prediction is a mapping and barely need iteration. Although there may exist some difference between the generated solution and the optimal solution, the DNN-based approach can achieve prediction with relatively high accuracy since it was proved to has capacity to approximate various complex mappings.

Another category is the heuristic algorithm methods based on computational intelligence techniques, including evolutionary programming like genetic algorithms \cite{Lai1997Improved} and swarm optimization \cite{Kumar2013Optimal}. There are two drawbacks of this kind of methods. First, both the optimality and feasibility of the solution is hard to guarantee. Second, the processing time it takes to find the optimal solution may also be substantial. Compared with these heuristic algorithms, the proposed DNN-based approach can achieve relatively accurate solution more quickly. 

The other category is the learning-based methods.
Some work focused on find an end-to-end approach to solve the OPF problem \cite{glavic2017reinforcement} based on reinforcement learning. However, the main drawback is that it is hard to guarantee the performance of the model as the model is trained without labeled data as supervisor, Meanwhile, it is difficult to guarantee both the accuracy and feasibility when the operation constraint is considered. 
In addition, the methods based on reinforcement learning are unable to leverage pre-existing knowledge about the mathematical form of the optimization problem \cite{misra2018learning}, \cite{amodei2016concrete}. Other work investigated how to integrate the traditional and the learning method to accelerate solving the OPF problem. For example, a principal component analysis (PCA) based approach was proposed in \cite{7448982} to extract the hidden relationships describing variables from operation data-sets in the OPF problems for reducing the number of variables to be solved, which can obtain solutions more efficiently. A statistical learning-based approach was proposed in \cite{Ng2018Statistical} to determine the active constraint set for the DC-OPF problem, which can reduce the scale of optimization whilst keep the highly accurate solutions. However, there are mainly two drawbacks of previous learning-based approaches: 1) For methods which use learning techniques to find the optimal solution without resorting the traditional solvers, they mainly focus on the prediction accuracy, which do not consider the dependency between variables of the OPF problem and the constraints.
Hence, these methods cannot keep the feasibility of the generated solution well; 2) For methods in which learning method is only used to assist the traditional solver, the machine models are applied to replace specific sub-steps of the traditional methods, whole algorithms still mainly depend on the traditional iteration method. As the times complexity may be high, these methods cannot be able to handle with large-scale power system. Unlike these method, the proposed approach can alleviate these two issues. On one hand, it does not depends on traditional OPF solver during the calculation, which can further reduce the time complexity. Also, we consider the constraints and integrate the post-processing into the proposed approach, which can guarantee both the accuracy and feasibility of generated solution. Apart from that, ~\cite{karg2018efficient} presented an approach to solve the constrained finite time optimal control problem based on deep learning, which mainly focuses on achieves high accuracy with less memory requirements. Despite such similarity, however, this paper mainly focuses on how to solve the OPF problem faster by levering the deep learning technique.

\section{The Optimal Power Flow Problem} \label{sec:OPF.review}
We review the formulations of general OPF and DC-OPF problems in this section. The objective of the OPF problem is to determine the operation states of the power system considering the relationship between the cost and supplied power of each generator as well as the constraints of system operation \cite{gomez2018electric}. We summarize the key notations used in this paper in Table~\ref{table0}.
\subsection{General OPF problem formulation} 
A general OPF problem can be formulated as follows:
\begin{eqnarray}
&{\min_{\left(\mathbf{x,u}\right)}}&f\left( \mathbf{x,u} \right) \nonumber\\
&\mathrm{s.t.\;\;}&g_p\left(\mathbf{x,u}  \right) =0,\ p=1,2,...,n_p, \\ 
&&h_m\left(\mathbf{x,u}  \right) \le 0,\ m=1,2,...,n_m, \nonumber
\label{equation1}
\end{eqnarray}
\noindent where $f( \cdot )$ is the objective function, which is usually designed according to certain economic criteria such as minimizing the production cost of of all generators. $\mathbf{x}$ and $\mathbf{u}$ are the state variables and the decision variables, respectively. The decision (state) variables on different buses are:
\begin{itemize}
\item The real and reactive powers (voltage magnitude and angle) at each load bus.
\item The real power generated and the voltage magnitude (reactive power generated and voltage angle) at each generation bus.
\item The voltage magnitude and angle (the real and reactive power generated) at the reference bus.
\end{itemize}
The equality constraints $g_p (p=1,2,...,n_p)$ correspond to the power flow balance equations. The inequality constraints $h_m (m=1,2,...,n_m)$ are constraints corresponding to operation limits for transmission lines. 
\begin{table}[!t]
	\caption{Summary of Notations.}
	\renewcommand{\arraystretch}{1.2}
	\centering
	\begin{tabular}{ll}
		\hline
		Notation & Definition\\
		\hline
		$P_{Gi}$ & The active power from the generator in the i$th$ bus\\
 		$\theta_{i}$ & The phase angle of the generator in the i$th$ bus\\
 		$\mathbf{B}$ & The admittance matrix\\
 		${x_{ij}}$ & The line reactance between the $i$th bus and the $j$th bus\\
		$N_{\mbox{bus}}$ & The number of bus\\
		$N_{\mbox{gen}}$ & The number of generator\\
		$N_{\mbox{load}}$ & The number of load\\
		$N_{\mbox{bran}}$ & The number of branches\\
		$N_{\mbox{hid}}$ & The number of hidden layer\\
		$N_{\mbox{neu}}$ & The number of neurons in each layer\\
		$lr$ & The learning rate in the training stage\\
		\hline
	\end{tabular}
	\label{table0}
\end{table}
\subsection{Problem formulation for DC-OPF}
As mentioned before, in this paper we focus on the DC-OPF problem. In the DC-OPF problem, there are only two types of variables, i.e., the generator outputs and the power phase angle of transmission lines. The problem is to  minimize the total generation cost subject to the generator operation limits, the power balance equation, and the transmission line capacity constraints \cite{823997}. It can be expressed as follows:
\begin{eqnarray}
&\min&\mathrm{\ }\sum_{i=1}^{N_{\mbox{gen}}}{C_i\left( P_{Gi} \right)}\nonumber\\
\label{equation2_1}
&\mathrm{s.t.}&\left\{ \begin{array}{l}
P_{Gi}^{\min}\le P_{Gi}\le P_{Gi}^{\max},\,\,i=1,2,...,N_{\mbox{bus}} \\
\mathbf{B}\cdot \mathbf{\theta }=\mathbf{P}_{\mathbf{G}}-\mathbf{P}_{\mathbf{D}} \\
\frac{1}{x_{ij}}\left( \theta _i-\theta _j \right) \le P_{ij}^{max},\,\,i,j=1,2,...,N_{\mbox{bus}} \\
\end{array} \right. \ 
\end{eqnarray}
\noindent where $N_{\mbox{bus}}$ is the number of buses and $N_{\mbox{gen}}$ is the number of generators. $P_{Gi}$ is the power output of the generator in the $i$th bus. $P_{Gi}^{\min}$ and $P_{Gi}^{\max}$ are the output limits of generators in the $i$th bus, respectively.  If there is no generator in the $i$the bus, $P_{Gi}^{\min}$ and $P_{Gi}^{\max}$ are set to be 0. $\mathbf{B}$ is an $N_{\mbox{bus}} \times N_{\mbox{bus}}$ admittance matrix. If the $i$the bus and the $j$the bus are adjacent buses, the the corresponding element in matrix $\mathbf{B}$ is the reciprocal of the line reactance $x_{ij}$, Otherwise the corresponding element is 0. In particular,
$\mathbf{B}$ can be expressed as:
\begin{equation}
\mathbf{B}=\left[ \begin{matrix}
y_{11}&		y_{12}&		...&		y_{1N_{\mbox{bus}} }\\
y_{21}&		y_{22}&		...&		y_{2N_{\mbox{bus}} }\\
...&		...&		...&		...\\
y_{ N_{\mbox{bus}}1}&		y_{ N_{\mbox{bus}}2 }&		...&		y_{ N_{\mbox{bus}}N_{\mbox{bus}} }\\
\end{matrix} \right] \
\label{equation}
\end{equation}
\noindent where 
\[ B_{ij} =
  \begin{cases}
    -y_{ij},      & \quad \text{if } i\neq j;\\
    \displaystyle\sum_{k=1,k\neq i}^{N_{\mbox{bus}}} y_{ik}, & \quad \text{if } i= j.
  \end{cases}
\]
Here $y_{ij}$ is the admittance of the branch $b_{ij}$. In addition, we have $y_{ij}>0$ if there is an branch connecting node $i$ and node $j$, and $y_{ij}=0$ otherwise \cite{867153}. As a consequence, $B_{ii}>0, \forall i \in \{1,2,\cdots, N_{\mbox{bus}}\}.$

$\mathbf{P}_{\mathbf{G}}$ is the bus power generation vector and $\mathbf{P}_{\mathbf{D}}$ is the bus consumption vector.  $\mathbf{\theta}$ is the phase angles vector. $\theta_i$ and $\theta_j$ are the phase angles at the $i$th bus and the $j$th bus, respectively.  $\frac{1}{x_{ij}}\left( \theta _i-\theta _j \right)$ represents the bus power injection from the $j$th bus and the $i$th bus. $P_{ij}^{\max}$ is the transmission limit from the $i$th bus to the $j$ bus. $C_i\left( P_{Gi} \right)$ is individual cost function for the generator in the $i$th bus. The cost function is derived from a heat rate curve, which gives the generator electric power output as a function of the thermal energy input rate times the fuel cost per thermal energy unit. It is commonly modeled as a quadratic function~\cite{260897}:
\begin{equation}
C_i\left( P_{Gi} \right) =\lambda _{1i}^2P_{Gi}+\lambda _{2i}P_{Gi}+\lambda _{3i},
\label{equation3}
\end{equation}
\noindent where $\lambda_{1i}$,$\lambda_{2i}$, and $\lambda_{3i}$ are the model parameters. As shown in the formulation of the DC-OPF problem, the first inequality constraint is related to the generator operation limit. The second equality constraint means the power flow balance equation, where the total power injection into one bus equals to the sum of load consumption on this bus and the power injection from this bus to the adjacent buses. The third constraint means that the power injection on each transmission line cannot exceed the transmission capacity of the line. The DC-OPF problem in (\ref{equation3}) is a quadratic programming problem. 

\section{{\textsf{DeepOPF}} for Solving DC-OPF} \label{sec:DeepOPF}
\begin{figure*}[!t]
	\centering
	\includegraphics[width = 0.75\textwidth]{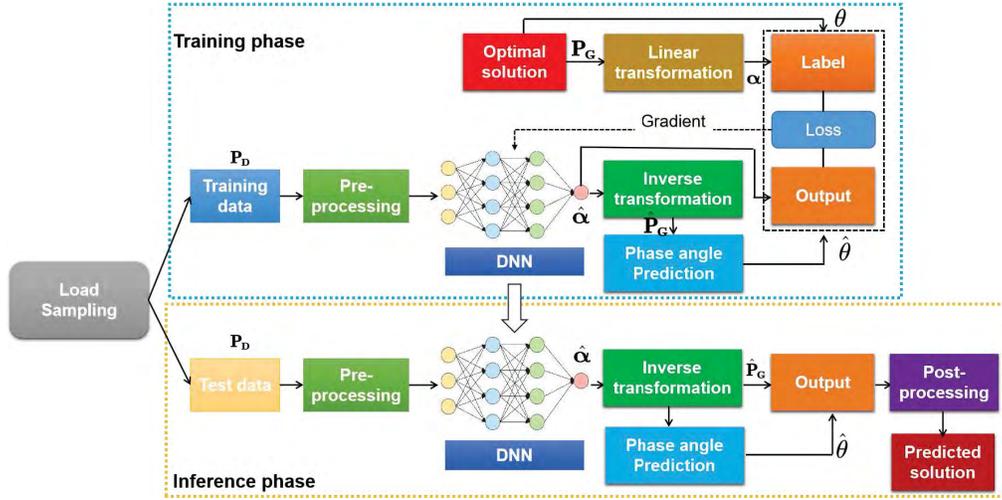}
	\caption{The framework of \textsf{DeepOPF}.}
	\label{fig3}
\end{figure*}

We propose \textsf{DeepOPF}, a DNN based framework for solving the DC-OPF problem. It integrates the structure of the DC-OPF problem into the design of the learning model. Specifically, we take the dependency between the variables to be solved into account and reformulate the constraints in the DC-OPF problem, which can reduce the number of the variables and transform the DC-OPF problem into a fitting problem. 

\subsection{Overview of \textsf{DeepOPF}}
The flowchart of the proposed framework is depicted in Fig. \ref{fig3}, which can be divided into training and inference stages. The key in \textsf{DeepOPF} is to construct a DNN model for mapping the load input into the active power and phase angle output. We outline the process of constructing and training the DNN model in the following.

\begin{mdframed}[skipabove=10pt, skipbelow=10pt, roundcorner=10pt, linewidth=0pt, backgroundcolor=gray!10] 

\indent First, we collect the training data and perform pre-processing. Specifically, we apply a uniform sampling method to generate the load $\mathbf{P}_{\mathbf{D}}$. We then obtain the optimal solution $\mathbf{P}_{\mathbf{G}}$ and $\theta$ for the corresponding DC-OPF problems as the ground-truth, by using a state-of-the-art solver~\cite{4282060}. After that, the training data will be normalized during the pre-possessing. The details are in Sec.~\ref{ssec:load.sampling.and.pre-preprocessing}. Note that the uniform sampling can address the over-fitting issue common in generic DNN approaches.

Second, we study an equivalent formation of DC-OPF to use a scaling factor vector $\mathbf{\hat{\alpha}}\in[0,1]$ to represent the active power $\mathbf{P}_{\mathbf{G}}$, normalizing the output value ranges which are known to facilitate training efficiency. Furthermore, we leverage on the fact that the admittance matrix (after removing the entries corresponding to the slack bus) is full rank to represent $\theta$ by $\mathbf{P}_{\mathbf{G}}$. Thus, it suffices to learn only the mapping from load inputs to the active powers instead of to both the active powers and the phase angles as in generic DNN approaches. This way, we can significantly reduce the mapping dimension, resulting in a much smaller size of the DNN model and amount of training data/time needed. The details are in Sec.~\ref{ssec:linear.transformation.and.dimension.reduction}. 

Third, we build a DNN model with $N_{\mbox{hid}}$ hidden layers with $N_{\mbox{neu}}$ on each layer based on the scale of the power system to solve the fitting problem of the scaling factor vector $\mathbf{\hat{\alpha}}$ after transformation in the second step. We train the DNN model by applying the data-driven stochastic gradient descent optimization algorithm to minimize a carefully-chosen loss function designed for DC-OPF problems. The details can be found in Sec.~\ref{ssec:DNN.and.loss.function}.

Fourth, we integrate post-processing into the \textsf{DeepOPF} to guarantee the feasibility of
the  solutions obtained by the DNN model. If the DNN model generates infeasible solution, \textsf{DeepOPF} will project the solution into the feasible region and return a feasible solution. The details are in Sec~\ref{ssec:post.processing}.
\end{mdframed}

\hspace{10pt}

In the inference stage, we directly apply \textsf{DeepOPF} to solve the DC-OPF problem with given test load inputs. We analyze the computational complexity of \textsf{DeepOPF} with a comparison to that of conventional approaches in Sec.~\ref{ssec:comp.complexity}. 

\subsection{load sampling and pre-processing}\label{ssec:load.sampling.and.pre-preprocessing}
As mentioned in Section I, we assume the load on each bus varies within a specific range around the default value independently. The load data is sampled within $[(1-x)*P_d, (1+x)*P_d]$ ($P_d$ is the default power load at individual bus $d$ and $x$ is the percentage of sample range like 10$\%$) uniformly at random. It is then fed into the traditional DC-OPF solver to generate the optimal solutions. Notice that here we adopt uniform sampling to avoid the over-fitting issue common in generic DNN approaches. However, this uniform mechanism may not be sufficient to guarantee enough sampling when the scale of the network become larger.  Markov chain Monte Carlo (MCMC) methods can be applied to this scenario, which sample based on a probability distribution. We can obtain a series of samples following the distribution that is used to construct the Markov chain. In our case, we can use MCMC to get samples near the boundary of the sampling space and obtain a dense sample set around the significant elements of the load vector.

As the magnitude for each dimension of the input and output may be different, The each dimension of training data will be normalized with the standard variance and mean of the corresponding dimension before training, which can make it easier for training. 

\subsection{Linear transformation and mapping dimension reduction} \label{ssec:linear.transformation.and.dimension.reduction}
There are inequality constraints related to $P_{Gi}$ on the problem. We first reformulated these inequality constraints through linear scaling as:
\begin{equation}
P_{Gi}\left( \alpha _i \right)=\alpha _i\cdot\left( P_{Gi}^{\max}-P_{Gi}^{\min} \right)+P_{Gi}^{\min} ,\ \alpha _i\in \left[ 0,1 \right] ,i=1,...,N_{\mbox{gen}},\ 
\label{equation4}
\end{equation}
\noindent where we recall that $N_{\mbox{gen}}$ is the number of generators. Thus, instead of predicting the generated power, we can predict the scaling factor $\alpha_i$ and obtain the value of $P_{Gi}$. There are two advantages of this approach. First, the output of the DNN is restricted within the range from 0 to 1 by using sigmoid function\cite{goodfellow2016deepma}. Thus, it can prevent the recovery of the predictions from violating the inequality constraints. Second, the prediction range becomes smaller, which makes it easier for the DNN model to learn the mapping between the load and $P_{Gi}$. It should be noted that one bus is set as the slack bus and used for balancing the mismatch between the total load and supply, thus the $P_{Gi}$ of the slack bus is obtained by subtracting output of the other buses from the total load.

As shown in Section II-B, there are two types of variables needed to be determined. Since the phase angle $\theta_i$ is the state variable depends on the decision variable $P_{Gi}$, and there exists a linear relationship between them (the second equality constraint in the DC-OPF problem), we can reduce the number of variables by representing $\theta_i$  with $P_{Gi}$. Thus, after obtaining $P_{Gi}, i=1,...,N_{\mbox{gen}}$, we can have bus power generation vector $\mathbf{P}_{\mathbf{G}}$, and obtain $\theta$ through $\mathbf{P}_{\mathbf{G}}$. Suppose we obtain the $(N_{\mbox{bus}}-1) \times (N_{\mbox{bus}}-1)$ matrix, $\mathbf{\tilde{B}}$ by eliminating corresponding row and column of the $n \times n$ admittance matrix $\mathbf{B}$ to the slack bus whose phase angle is usually set to zero. The following lemma states a useful property of the admittance matrix B pretty well-known in the literature; see e.g., \cite{823997}, \cite{7962226}.
\begin{lemma}
	The matrix $\mathbf{\tilde{B}}$ is a full-rank matrix.
	\label{lemma1}
\end{lemma}

The result is rather well known in the literature of power system; see e.g., \cite{823997}, \cite{7962226}. We provide a brief proof in Appendix \ref{apx:proof_of_lemma_1} for completeness.

According to Lemma~\ref{lemma1}, the matrix $\mathbf{\tilde{B}}$ is invertible. Thus, we can express the phase angles of all bus except the phase angle of the slack bus as following:
\begin{equation}
\tilde{\theta}=\left( \mathbf{\tilde{B}} \right) ^{-1}\left( \mathbf{\tilde{P}}_{\mathbf{G}}-\mathbf{\tilde{P}}_{\mathbf{D}} \right),
\label{equation5} 
\end{equation}
\noindent where $\mathbf{\tilde{P}}_{\mathbf{G}}$ and $\mathbf{\tilde{P}}_{\mathbf{D}}$ stand for the $(N_{\mbox{bus}}-1)$-dimension output and load vectors for buses excluding the slack bus, respectively. As the phase angle of the slack bus is fixed, the DNN model does not need to predict it. After the transformation, we can obtain the generated phase angle vector $\hat{\theta}$ by inserting zero representing the phase angle for the slack bus into $\tilde{\theta}$. There are mainly two advantages of this transformation. On one hand, we use the property of the admittance matrix to reduce the number of variables, which can further reduce the size of our DNN model and the amount of training data/time needed. It means the usage of the structure of the DC-OPF problem can lead to more efficient design of the DNN model.
On the other hand, the operations in the transformation is differentiable with respect to the generated $\mathbf{\hat{P}_G}$, which makes it convenient to introduce error related to $\theta$ in the loss function.

For the inequality constraints related to the transmission on each line, in order to satisfy these requirements, we can represent these constraints by $\mathbf{P}_{\mathbf{G}}$, and add a penalty term related to these constraints into the loss function to guide the training of the DNN model. It should be noted that the inequality constraints related to the transmission line between adjacent buses is symmetric, thus the inequality can be reformulated as following:
\begin{equation}
-1\le \frac{1}{P_{ij}^{\max}\cdot x_{ij}}\cdot \left( \theta _i-\theta _j \right) \le 1,,\,\,i,j=1,2,...,N_{\mbox{bus}}.
\label{equation6}
\end{equation}
The lower bound and upper bound of the inequality for each transmission line can be -1 and 1 after transformation, respectively. Thus, we can introduce a penalty function in the loss function of the DNN model, which can make the generated solution satisfy the constraints better. The applied penalty function is:
\begin{equation}
p\left( x \right) = x^2-1.
\label{equation7}
\end{equation}
To integrate this penalty term in the loss function, we can express the (\ref{equation6}) with $\mathbf{P}_{\mathbf{D}}$. We first introduce an $n_a \times n$ matrix $\mathbf{A}$, where $n_a$ is the number of adjacent buses. Each row in $A$ corresponds to an adjacent bus pair. Given any the adjacent bus pair $(i,j)$, we assume the power flow is from the $i$th bus to the $j$th bus. Thus, the elements, $a_{i}$ and $a_{j}$, are the corresponding entries of the matrix $\mathbf{A}$ defined as:
\begin{equation}
a_i=\frac{1}{P_{ij}^{\max}\cdot x_{ij}}\ \mathrm{and\ }a_j=-\frac{1}{P_{ij}^{\max}\cdot x_{ij}}.
\label{equation8}
\end{equation}
Based on (\ref{equation5}), (\ref{equation6}) and (\ref{equation8}) can be expressed as:
\begin{equation}
-1\le \left(\mathbf{A}\hat{\theta} \right)_k\le 1, k=1,...,n_a,
\label{equation9}
\end{equation}
\noindent where $\left(\mathbf{A}\hat{\theta} \right)_k$ represents the $k$th element of $\mathbf{A}\hat{\theta}$. Thus, the phase angle vector $\hat{\theta}$ is obtained through $\mathbf{P}_{\mathbf{G}}$ and $\bar{\theta}$. We can then calculate the penalty value for $\left(\mathbf{A}\hat{\theta} \right)_k$, and add the average penalty value into the loss function for training.

The transformation is to make the training easier by considering the structure of the DC-OPF problem. For one thing, we take dependency between the active power and the phase angle into account and represent the phase angle with the active power, which can reduce the number of variables to be generated. For the other, the constraints of the DC-OPF are more easier to be dealt with. After transformation, the constraints related to the generator output and the power balance can be well satisfy. Meanwhile, the penalty is introduced to represent the constraint on each transmission line. Thus, solving the DC-OPF problem changes to the fitting problem with respected to $P_{Gi}\left( \alpha _i \right)$ plus a penalty term, which means if the DNN model can approximate the mapping between the input parameter and the output, the highly accurate generated solution for the DC-OPF problem can be obtained immediately without resorting to traditional iteration based solvers. 

\subsection{The DNN model} \label{ssec:DNN.and.loss.function}
The core of \textsf{DeepOPF} is the DNN model applied to approximate the mapping between the load and power output of the generators. The DNN model is established based on the multi-layer feed-forward neural network structure, which consists of a typical three-level network architecture: an input layer, several hidden layers, and an output layer. More specifically, the applied DNN model is defined as:
\begin{eqnarray*}
	&h_0&=\mathbf{\tilde{P}}_{\mathbf{D}},\\
	&h_i&=\sigma \left( W_ih_{i-1}+b_{i-1} \right),\\ 
	&\hat{\alpha}&=\sigma '\left( w_oh_L+b_o \right), 
\end{eqnarray*}
where $h_0$ denotes the input vector of the network, $h_i$ is the output vector of the $i$th
hidden layer, $h_L$ is the output vector (of the output layer), and $\hat{\alpha}$ is the generated scaling factor vector for the generators. 

\subsubsection{The architecture}
In the DNN model, $h_0$ represents the normalized load data, which is the inputs of the network. After that, features are learned from the input vector $h_0$ by several fully connected hidden layers. The $i$th hidden layer models the interactions between features by introducing a connection weight matrix $W_i$ and a bias vector $b_i$. Activation function $\sigma(\cdot)$ further introduces non-linearity into the hidden layers. In our DNN model, we adopt Rectified Linear Unit (ReLU) as the activation function of the hidden layers, which can be helpful for accelerating the convergence and alleviate the vanishing gradient problem \cite{krizhevsky2012imagenet}. At last, the Sigmoid function $\sigma '\left( x \right) =\frac{1}{1+e^{-x}}$  is applied as activation function of the output layer to project the outputs of the network to $(0, 1)$. 

For different power networks (as IEEE test cases), we tailor the DNN model by educated guesses and iterative tuning, which is by far the common practice in generic DNN approaches in various engineering domains. The number of hidden layers $N_{\mbox{hid}}$ and neurons on each hidden $N_{\mbox{neu}}$ of DNN is mainly designed according to the scale of the power system. For example, the mapping for larger scale power system is more difficult to obtain as there are more mutual dependent variables to predict, thus DNN model with larger $N_{\mbox{hid}}$ and $N_{\mbox{neu}}$ will be used for larger power system. 

An example of the DNN architecture is shown Fig.~\ref{fig4}, which we construct for the IEEE Case30 test case in the numerical experiments in Sec.~\ref{sec:simulations}.

\subsubsection{The loss function}
After constructing the DNN model, we need to design the corresponding loss function to guide the training. For each item in the training data set, the loss function consists of two parts: the difference between the generated solution and the reference solution obtained from solvers and the value of the penalty function upon solutions being infeasible. Since there exists a linear correspondence between $\mathbf{P}_{\mathbf{G}}$ and $\theta$, there is no need to introduce the loss term of the phase angles. The difference between the generated solution and the actual solution of $P_{Gi}$ is expressed by the sum of mean square error between each element in the generated scaling factors $\hat{\alpha}_i$ and the actual scaling factors ${\alpha}_i$ in the optimal solutions:
\begin{eqnarray}
\mathcal{L}_{PG}=\frac{1}{N_{\mbox{gen}}}\sum_{i=1}^{N_{\mbox{gen}}}{\left( \hat{\alpha}_i-\alpha _i \right) ^2},
\label{equation10}
\end{eqnarray}
where $N_{\mbox{bus}}$ represents the number of generators. 
The penalty term capturing the feasibility of the generated solutions can be expressed as:
\begin{eqnarray}
\mathcal{L}_{pen}=\frac{1}{n_a}\sum_{k=1}^{n_a}{p\left( \left(\mathbf{A}\hat{\theta}\right)_k \right)},
\label{equation12}
\end{eqnarray}
\noindent where we recall that $n_a$ is the number of the adjacent buses, $p(\cdot)$ is the penalty function defined in $\left(\ref{equation7}\right)$, and $\left(\mathbf{A}\hat{\theta}\right)_k $ represents the $k$th element in the vector $\mathbf{A}\hat{\theta}$ ($\hat{\theta}$ is the generated phase angle vector by (\ref{equation5})). The total loss can be expressed as the weighted summation of the two parts:
\begin{eqnarray}
\mathcal{L}_{total}=w_1\cdot \mathcal{L}_{PG}+w_2\cdot \mathcal{L}_{pen},
\label{equation13}
\end{eqnarray}
\noindent where $w_1$ and $w_2$ are positive weighting factors, which are used to balance the influence of each term in the training phase. As the objective of the designed loss function is to find the relatively accurate $\mathbf{P}_{\mathbf{G}}$, the first term is with the highest priority in the loss function as it has much more influence on the recovery of the $\mathbf{P}_{\mathbf{G}}$. The other loss term, which is with respect to the penalty related to the transmission line, is regarded with lower priority in the training process. 

\subsubsection{The training process}
In general, the training processing can be regarded as minimizing the average value of loss function with the given training data by tuning the parameters of the DNN model as follows:
\begin{equation}
    \min_{W_i, b_i }\frac{1}{N_{\mbox{train}}}\sum_{k=1}^{N_{\mbox{train}}}{\mathcal{L}_{total,k}}
\end{equation}
\noindent where $W_i$ and $b_i$ represent the connection weight matrix and vector for layer $i$.
$N_{\mbox{train}}$ is the amount of training data and $\mathcal{L}_{total,k}$ is the loss of the $k$th item in the training. 
We apply the widely-used optimization technique in the deep learning, stochastic gradient descent (SGD) \cite{goodfellow2016deepma},  in the training stage, which is effective for large-scale dataset and can economize on the computational cost at every iteration by choosing a subset of summation functions at every step. We outline the process of SGD in the following Algorithm~\ref{alg:SGD}.
\begin{algorithm}
	\caption{SGD}\label{alg:SGD}
	\begin{algorithmic}[1]
		\State Initialize the DNN model parameters ($W_i$, $b_i$), and the learning rate $lr$.
		\State Set the number of batch size and total number of epochs T (an epoch means the round that the all batches have been trained).
		\State Set the start round index $t=1$
		\Repeat  
		\State Randomly shuffle examples in the training set;
		\For{Each batch in the training data set}                    
			\State Calculate the average loss for each batch, $\mathcal{L}^t$ in the feed-forward manner;
		\State Update the parameters $W_i$ and $b_i$ in the back propagation manner with the gradient as follows: 
		\begin{equation*}
		W_{i}^{t+1}=W_{i}^{t}-lr\cdot \nabla \mathcal{L}^t\left( W_{i}^{t} \right)
		\end{equation*}
		\begin{equation*}
		b_{i}^{t+1}=b_{i}^{t}-lr\cdot \nabla \mathcal{L}^t\left( b_{i}^{t} \right)
		\end{equation*}
		where $\nabla \mathcal{L}^t\left( W_{i}^{t} \right)$ and $\nabla \mathcal{L}^t\left( b_{i}^{t} \right)$ represent the gradient with respect to the parameters $W_i$ and $b_i$ in the $t$th round, respectively;   
		\EndFor
		\State t=t+1
		\Until {$t = T $}  
	\end{algorithmic}
\end{algorithm}

\subsection{Post-processing} \label{ssec:post.processing}
The proposed approach may encounter a situation in which the balanced amount of electricity exceeds the feasible capacity range of the slack bus. To address this issue and ensure the integrity of the algorithm, after the generated solution $\mathbf{\hat{P}_G}$ is obtained, the approach needs to conduct post-processing to guarantee the feasibility of the solution in case the DNN's generated solution is infeasible. 

As discussed before, the constraints in the DC-OPF problem are (closed) linear constraints. Thus, the post-processing can be regarded as to project the initial generated solution into the polyhedron that is the intersection of a finite number of closed linear constraints, which can be formulated as follows:
\begin{equation}
{\min}\ \lVert \mathbf{\hat{P}_G}-\mathbf{u} \rVert ^2
\mathrm{\ \ s.t.\ \ }\mathbf{u}\in C_1\cap C_2... \cap C_d,
\label{equation14}
\end{equation}
\noindent where the convex sets $C_1,C_2, ... , C_d$ denote the constraints in the DC-OPF problem. By solving the problem in  \eqref{equation14}, we can find the feasible solution closest to the generated solution $\mathbf{\hat{P}_G}$. The reason that we want to find the feasible one that is closest to the pseudo-optimal solution lies in that, we want to provides the feasible one which has the smallest variation from $\hat{P}_G$, and expecting the value of the cost function is also close to the optimum. The above problem is a convex quadratic programming problem and can be solved by the fast dual proximal gradient algorithm proposed in \cite{beck2014fast}, with a convergence rate of primal sequence being of the order $\mathcal{O}\left(1/\tau\right)$, where $\tau$ is the iteration steps. With the post-processing, the proposed approach can guarantee the feasibility of the generated solution.The numerical experiments on IEEE test cases in Sec.~\ref{sec:simulations} show that  \textsf{DeepOPF} generates feasible solutions for all DC-OPF instances; thus in practice, maybe only few instance will involve the post-processing process. 

\subsection{Computational complexity} \label{ssec:comp.complexity}
As mentioned before, the computational complexity of the traditional iteration approach is related to the scale of the DC-OPF problem. For example, the computational complexity of interior point method based approach for the convex quadratic programming is $\mathcal{O} \left(L^2  n^4\right)$ measured as the number of arithmetic operations \cite{ye1989extension}, where $L$ is the number of input bits and $n$ is the number of  variables. 

For our proposed approach, the computational complexity mainly consists of two parts: the calculation as the input data passing through the DNN model and the post-processing. For the post-processing, its computational complexity can be negligible in the practical usage as the DNN model barely generate infeasible solution. Thus, the computational complexity of the proposed approach is approximately determined by the calculation with respect to the DNN model. It can be evaluated by the scale of the network \cite{7299173}. 

Recall that the input and the output of the DNN model in \textsf{DeepOPF} are $N_{\mbox{in}}$ and $N_{\mbox{out}}$ dimensions, respectively, and the DNN model has $N_{\mbox{hid}}$ hidden layers and each hidden layer has $N_n$ neurons. Once we finish training the DNN model, the complexity of generating solutions by using \textsf{DeepOPF} is characterized in the following proposition.
\begin{proposition}
The computational complexity (measured as the number of arithmetic operations) to generate a solution to the DC-OPF problem by using \textsf{DeepOPF} is given by 
\begin{equation}
T=N_{\mbox{in}} N_{\mbox{n}}+ (N_{\mbox{hid}}-1) N_{\mbox{n}}^2 + N_{\mbox{out}} N_{\mbox{n}},
\label{complexity1}
\end{equation}
which is $\mathcal{O}\left(N_{\mbox{hid}} N_{\mbox{n}}^2\right)$. 
\end{proposition}
For example, as shown in Table \ref{table2}, for the parameters associated with IEEE 300 case, $T = 81920$. As for the given topology, the scale of the DNN model is fixed. It means the computational complexity $T$ is a constant. Thus, unlike the traditional methods, the proposed approach can find the mapping between the given parameter and the optimal solution without iteration process when the training is finished. We also conduct experiment to compare the actual processing time between the traditional and the proposed methods, where the numerical results show the proposed approach can obtain the solution in a much shorter time.

\begin{figure}[!ht]
	\centering
	\includegraphics[width = \columnwidth]{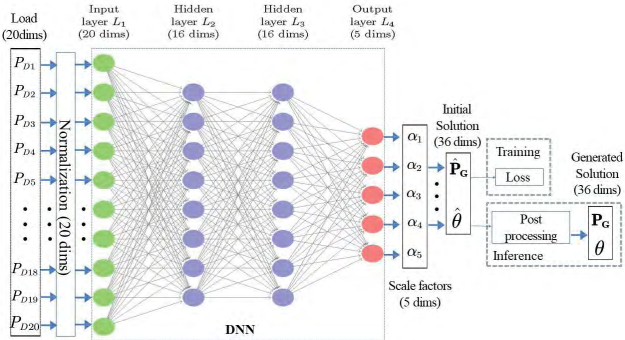}
	\caption{The detail architecture of DNN model for IEEE case30.}
	\label{fig4}
\end{figure}
\begin{table}[!t]
	\centering
	\caption{Parameters for test cases.}
	\renewcommand{\arraystretch}{1.0}
	\begin{threeparttable}
		\begin{tabular}{c|c|c|c|c|c|c|c}
			\toprule
			\hline
			Case & \tabincell{c}{$N_{\mbox{bus}}$} & \tabincell{c}{$N_{\mbox{gen}}$} & \tabincell{c}{$N_{\mbox{load}}$} &\tabincell{c}{$N_{\mbox{bran}}$} &\tabincell{c}{$N_{\mbox{hid}}$} &\tabincell{c}{$N_{\mbox{neu}}$} &\tabincell{c}{$lr$} \\
			\hline
			\tabincell{c}{IEEE \\Case30} & 30 & 6 & 20 &41 & 2&16&1e-3\\
			\hline
			\tabincell{c}{IEEE \\Case57} & 57 & 7 & 42 &80&4&32&1e-3\\
			\hline
			\tabincell{c}{IEEE \\Case118} & 118 & 54 & 99 &186&6&64&1e-3\\
			\hline
			\tabincell{c}{IEEE \\Case300} & 300 & 69 & 199 &411&6&128&1e-3\\
			\hline
			\bottomrule
		\end{tabular}
		\begin{tablenotes}
			\footnotesize
			\item[*] The number of load buses is calculated based on the default load on each bus. If the default load for active power on the bus does not equal to zero, the bus is considered as a load bus and vice versa.
			\item[*] The values for these parameters are not unique. Different combinations of the parameters may achieve similar performance.
		\end{tablenotes}
	\end{threeparttable}
	\label{table2}
\end{table}
\begin{table*}[!t]
	\centering
	\caption{PERFORMANCE EVALUATION OF THE PROPOSED APPROACH.}
	\renewcommand{\arraystretch}{1.4}
	\begin{threeparttable}
			\begin{tabular}{c|c|c|cc|cc|c}
			\toprule
			\hline
			\multirow{2}{*}{Case} &
			\multirow{2}{*}{\tabincell{c}{Number of \\Variables}} &
			\multirow{2}{*}{\tabincell{c}{Feasible solution's percentage\\ without post-processing (\%)}} &
			\multicolumn{2}{c|}{\tabincell{c}{Average cost \\(\$/hr)}} &
			\multicolumn{2}{c|}{\tabincell{c}{Running time \\ (millisecond)}} &
			\multicolumn{1}{c}{Speedup} \\
			\cline{4-7}
			&&&\textsf{DeepOPF}&Ref.&\textsf{DeepOPF}&Ref.&\\
			\hline
			\tabincell{c}{IEEE \\Case30} &36& 100 & 589 & 588 &  0.19& 21 & $\times$110\\
			\hline
			\tabincell{c}{IEEE \\Case57} &64& 100 & 42750  & 42667 &0.22& 27 & $\times$122\\
			\hline
			\tabincell{c}{IEEE \\Case118}&172& 100 &133776  & 131311 &0.29 &44& $\times$151\\
			\hline
			\tabincell{c}{IEEE \\Case300}&369& 100 &706602 &706338 &0.37& 50  & $\times$ 135\\
			\hline
			\bottomrule
		\end{tabular}
		\begin{tablenotes}
		\item[*] Running time is the average computation time of all test instances.
		\end{tablenotes}
		\end{threeparttable}
	\label{table4}
\end{table*} 
\begin{figure*}[!t]
	\centering
	\subfigure [] {\includegraphics[width = 0.48\textwidth]{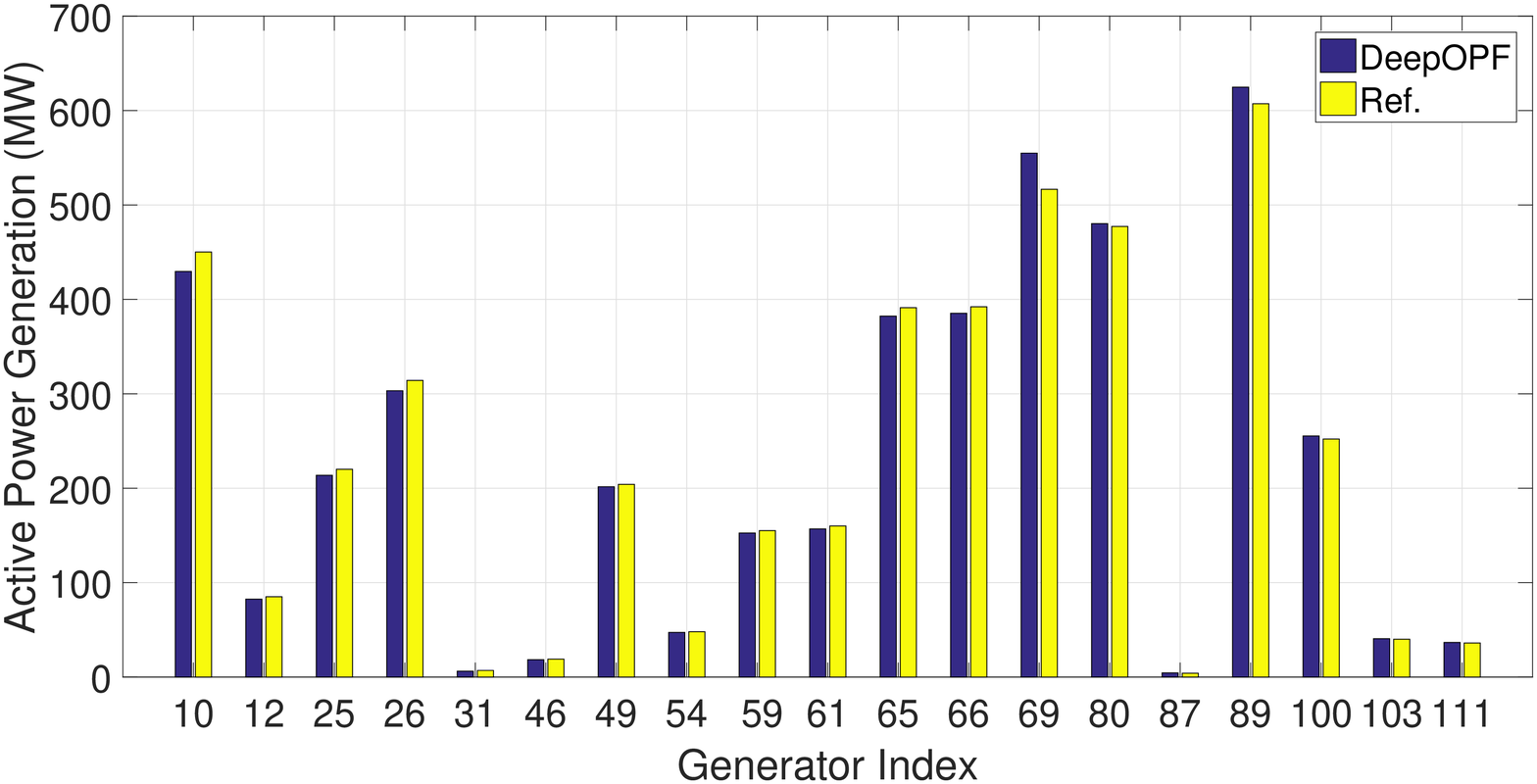}}
	\subfigure [] {\includegraphics[width = 0.48\textwidth]{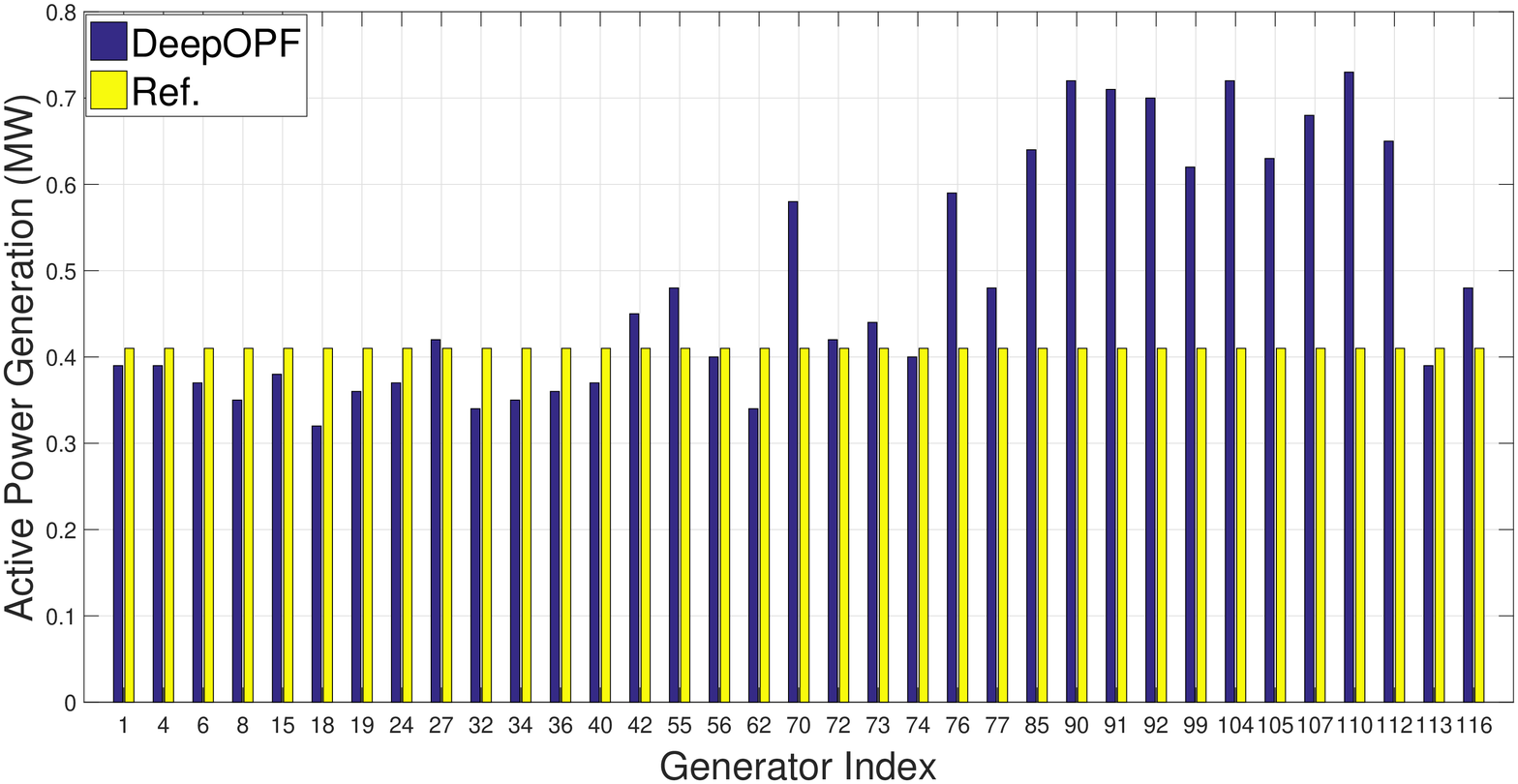}}
	\caption{(a)-(b) represent the comparisons of generated solution and the optimal solution for IEEE case-118 with the total load equals to 4393.53 MW . The total cost for the generated solution and the optimal solution is 132418 (\$/hr) and 131967 (\$/hr), respectively.}
	
	\label{fig2}
\end{figure*}

\section{Numerical Experiments}\label{sec:simulations}
\subsection{Experiment setup}
\subsubsection{Simulation environment} The experiments are conducted in Ubuntu 18.04 on the six-core (i5-8500@3.00G Hz) CPU workstation and 8GB RAM.
\subsubsection{Test case}
The proposed approach is tested with four IEEE standard cases \cite{tpcwTrey4}: the IEEE 30-bus power system, IEEE 57-bus power system, the IEEE 118-bus test system and the IEEE 300-bus system, respectively, which includes the small-scale, medium-scale and large-scale power system network for the DC-OPF problem. The related parameters for the test cases are shown in Table \ref{table2}. The illustrations of the topology for IEEE 30-bus system and 57-bus system are shown in Appendix as examples. The illustrations for the IEEE 118-bus and IEEE 300-bus cases can be found in \cite{4077121} and \cite{tpcwTrey6}. 

\subsubsection{Training data}
In the training stage, the load data is sampled within $[90\%, 110\%]$ of the default load on each load uniformly at random. After that, the solution for the DC-OPF problem provided by the pypower \cite{tpcwTrey1} is regarded as ground-truth. Pypower is based on the traditional interior point method \cite{4282060}. Taking the sampling range and different scale of the cases, the corresponding amount of training data for different cases is empirically determined in the simulation as follows: 10000 training data for IEEE Case30, 25000 training data for IEEE Case57 and IEEE Case118, 50000 training data for IEEE Case300. For each test case, the amount of test data is 10000.

\subsubsection{The implementation of the DNN model}
We design the DNN model based on Pytorch platform and apply the stochastic gradient descent method \cite{goodfellow2016deepma} to train the neural network. In addition, the epoch is set to 200 and the batch size is 64. Based on the range of each loss obtained from some preliminary experiments, the value of weighting factors $w_1$ and $w_2$ are set to 1 and 0.00001 term empirically. Meanwhile, other parameters includes the number of hidden layers, the number of neurons in each layer and the learning rate for each test case are also shown in Table. \ref{table2}. For better understanding, the detail architecture of the DNN model with input and post-processing for IEEE Case30 is shown in Fig. \ref{fig4}, which includes the dimension of the input and output as well as the structure (e.g., the number of neurons on each layer and the corresponding activation function). 

\subsection{Performance evaluation}
The simulation results of the proposed approach for test cases are shown in the Table \ref{table4}. We can see from the Table. \ref{table4} the percentage of the feasible generated solution is 100\% before post-processing, which indicates the proposed neural network model can keep the feasibility of the generated solution well without resorting to the post-process step. As the load of each node are assumed varying within a specific range, and the load for the training data are sampled in this range uniformly at random, the neural network model can find the mapping between the load in the specific range and the corresponding solution for the DC-OPF problem. It should be noted that the proposed neural network can learn the mapping for any varying range as long as there are enough sample data for training. Thus, the neural network model barely generates the infeasible solution, which means the proposed approach can mostly find the feasible solution through mapping. Even if it obtains the infeasible solution, the model can resort to the post-processing to adjust the infeasible generated solution into the feasible one. In addition, the difference between the cost with generated solution and that of the reference solution is shown in the Table. \ref{table4}. The difference can be negligible, which means each dimension of generated solution has high accuracy when compared to that of the optimal solution. 

To verify this, we show the comparisons between the generated solution and the optimal solution for all generators under IEEE case-118 with the a given total load in the Fig. 3 as an example. For better illustration, we divide the generators into two groups based on their optimal output value. Fig. 3(a) shows the comparisons of the generators with large output while Fig. 3(b) shows that of the generators with small output. During the training process, our DNN model tries to learn the relative relationship between the load and the optimal output, i.e., the proportion of electricity generated by each generator. In Fig. 3, the generator index means the index of the bus which the generator belongs to. We can observe from Fig. 3(a) that for prediction of the generators large output, the \textsf{DeepOPF} approach can not only describe the relative relation between each dimension in the optimal solution, but also obtain prediction with small difference. Usually, the total cost is mainly determined by the generators with larger output. While for the generators with small output, the \textsf{DeepOPF} can also obtain a relative accurate prediction as shown in Fig. 3(b). In our example, the cost of the generators with small output usually only accounts about 1\% percentage of the total cost, it will not have significant influence on the total cost although there exists difference. Thus, the proposed \textsf{DeepOPF} can effectively achieve the high accuracy solution without resorting to the traditional DC-OPF solver.

In addition, we show the histogram of prediction errors of different generators in the test stage for IEEE 30-bus case as an example in Appendix \ref{apx:histogram2}. As seen, the range of prediction error is acceptable for the practical operation, which can demonstrate the effectiveness of the DeepOPF for multi-dimension prediction. More discussion can be found in Appendix \ref{apx:histogram2}.

Apart from that, we can see that compared with the traditional DC-OPF solver, our DeepOPF approach can speed up the computing time by two order of magnitude. As mentioned before, given the topology of the power system, solving the OPF problem means to find the mapping between the load and decision variables of the generators. The proposed model can achieve high prediction accuracy much faster than the traditional iteration based solution. In addition, we provide the training time consumption (sec./epoch) as follows: case30 (0.3), case57 (3.6), case118 (10.4) and case300 (19.0). Recall that 200 epoches can achieve acceptable predicted solutions as shown in the simulation. As the OPF problem has to be solved frequently and repeatedly, the training time is negligible for the long-term DC-OPF application.

\subsection{The benefit of multi-layer structure}
We also carry out comparative experiments to show the benefits brought by the scale of network on the performance for the DC-OPF problem. More specifically, we design three variants of \textsf{DeepOPF} for IEEE case-57 with different depths, which are listed as follows:
\begin{itemize}
\item \textsf{DeepOPF}-V1: A simple network without hidden layer
\item \textsf{DeepOPF}-V2: A simple network with one hidden layer
\item \textsf{DeepOPF}-V3: A simple network with three hidden layers
\end{itemize}
where \textsf{DeepOPF}-V1 is the simplest network which has no hidden layers while \textsf{DeepOPF}-V3 has three hidden layers among the three variants. Comparative experiments follow the same experimental settings above and the results of all variants in terms of average cost (\$/hr) as well as percentage (\%), the relative increase of the cost compared to the optimal solution, and the running time (sec.) are shown in Table. \ref{table5}.
\begin{table}[!t]
	\caption{Performance comparisons of all variants.}
	\renewcommand{\arraystretch}{1.5}
	\centering
	\begin{tabular}{lcccc}
		\hline
		Variant & \tabincell{c}{Average cost \\(\$/hr)}& \tabincell{c}{$\Delta$ cost \\(\$/hr)/(\%)}& \multicolumn{1}{c}{\tabincell{c}{Running time \\(millisecond)}} \\
		\hline
		\textsf{DeepOPF}-V1 & 42854&+187/+0.4&0.12\\
		\hline
		\textsf{DeepOPF}-V2 & 42825&+158/+0.4&0.14\\
		\hline
 		\textsf{DeepOPF}-V3 & 42750&+83/+0.2&0.17\\
		\hline
	\end{tabular}
	\label{table5}
\end{table}
We can observe that increasing the depth of network does enhance the recommendation performance as \textsf{DeepOPF}-V3 outperforms other variants on average cost. A deeper network can model more interactions between features by introducing more network parameters while it takes a long time to train the model and is more sensitive to hyper parameters as well.

\section{Conclusion}\label{sec:conclusion}
Solving DC-OPF optimally and efficiently is crucial for reliable and cost-effective power system operation. In this paper, we develop \textsf{DeepOPF} as a DNN approach to generate feasible solutions for DC-OPF with negligible optimality loss. \textsf{DeepOPF} is inspired by the observation that solving DC-OPF for a given power network is equivalent to learning a high-dimensional mapping between the load inputs and the dispatch and transmission decisions. Simulation results show that \textsf{DeepOPF} scales well in the problem size and speeds up the computing time by two order of magnitude as compared to conventional of using modern convex solvers. These observations suggest two salient features particularly appealing for solving the (large-scale) security-constrained DC-OPF problem, which is central to secure power system operation with contingency in consideration. A compelling future direction is to apply \textsf{DeepOPF} to solve the non-convex AC-OPF problem.

\bibliographystyle{ACM-Reference-Format}
\bibliography{ref}

\appendix      

\begin{figure*}[ht!]
    \centering
	\subfigure [] {\includegraphics[width = 0.3\textwidth]{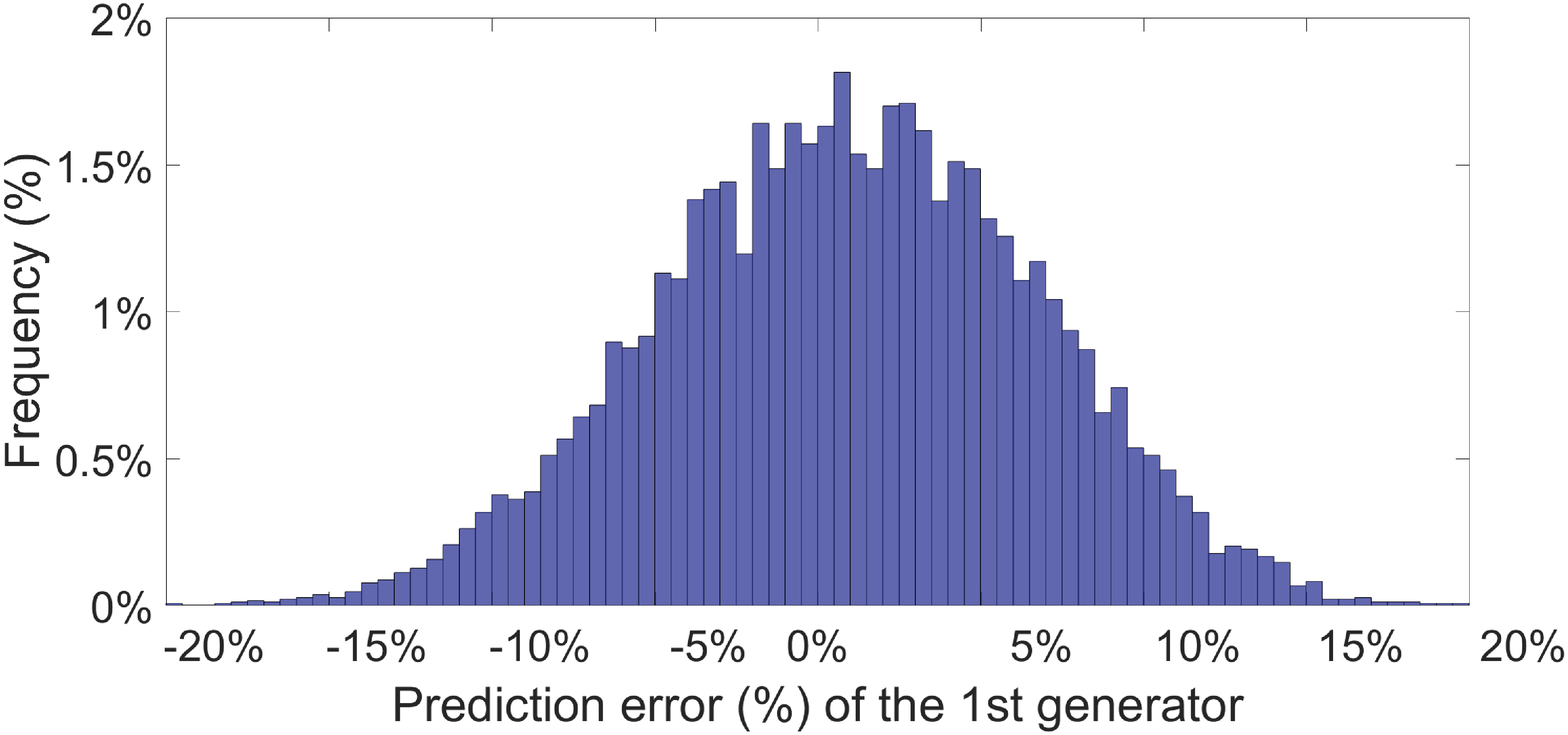}}
	\subfigure [] {\includegraphics[width = 0.3\textwidth]{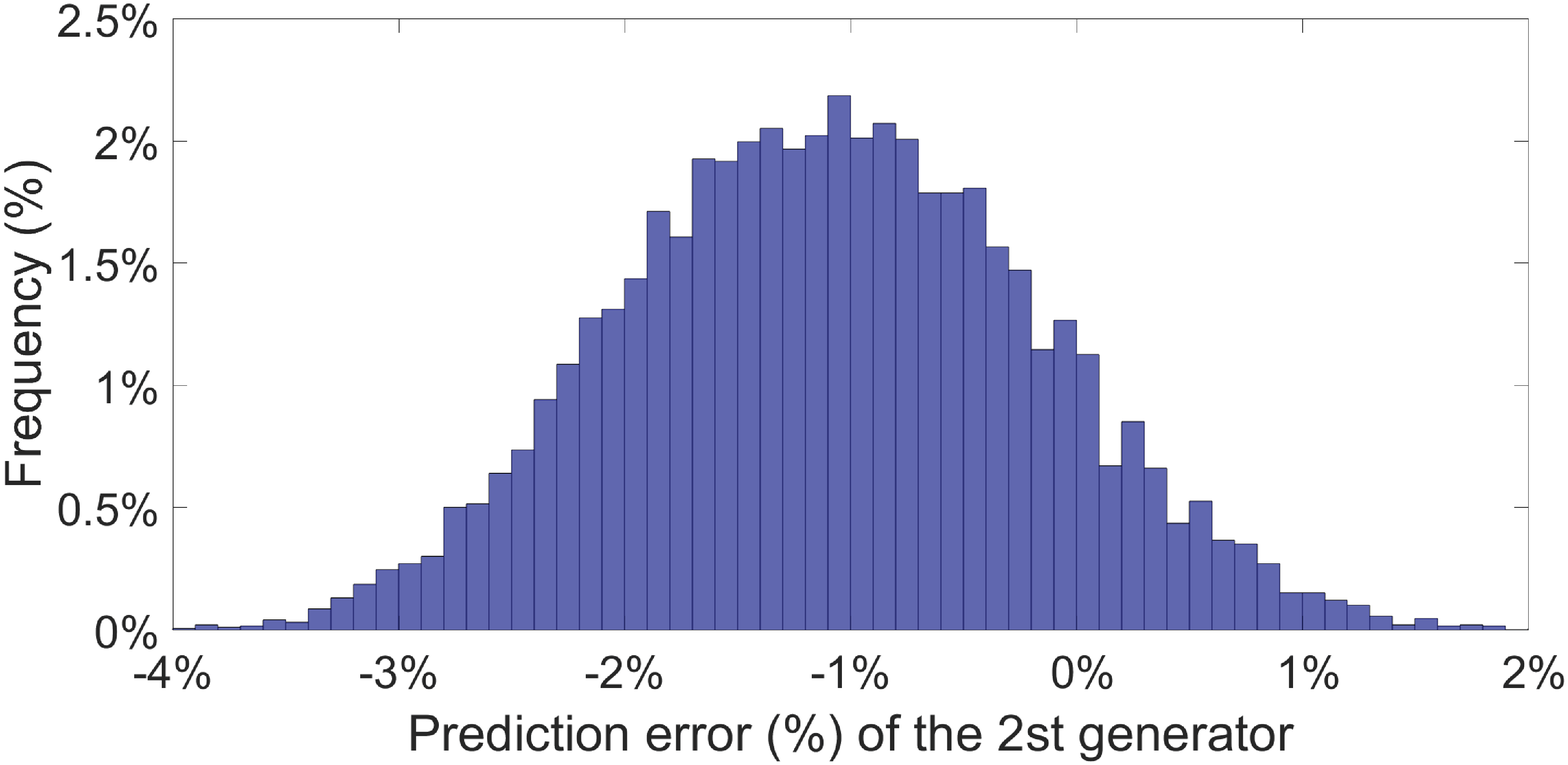}}
	\subfigure [] {\includegraphics[width = 0.3\textwidth]{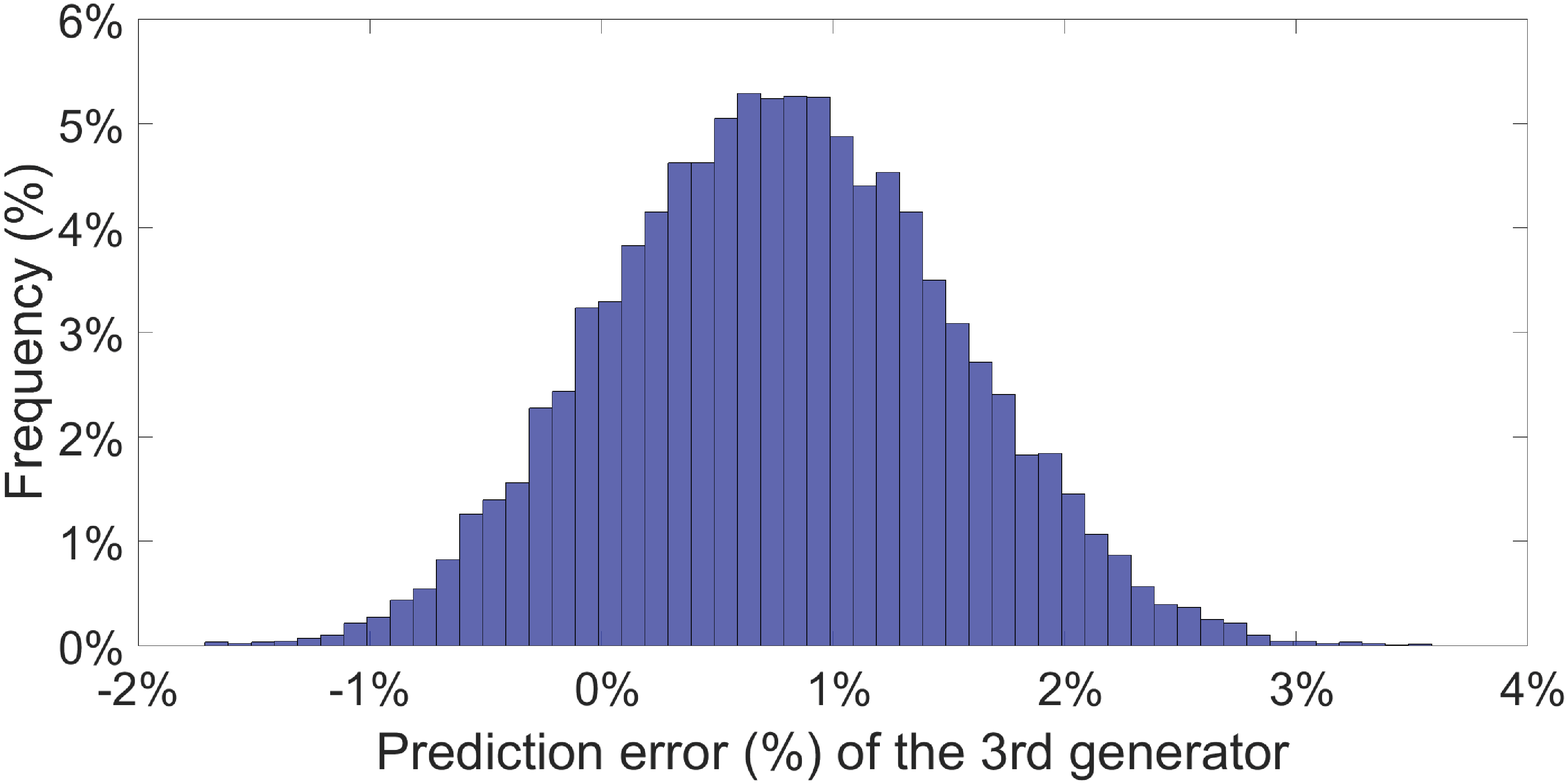}}
	\subfigure [] {\includegraphics[width = 0.3\textwidth]{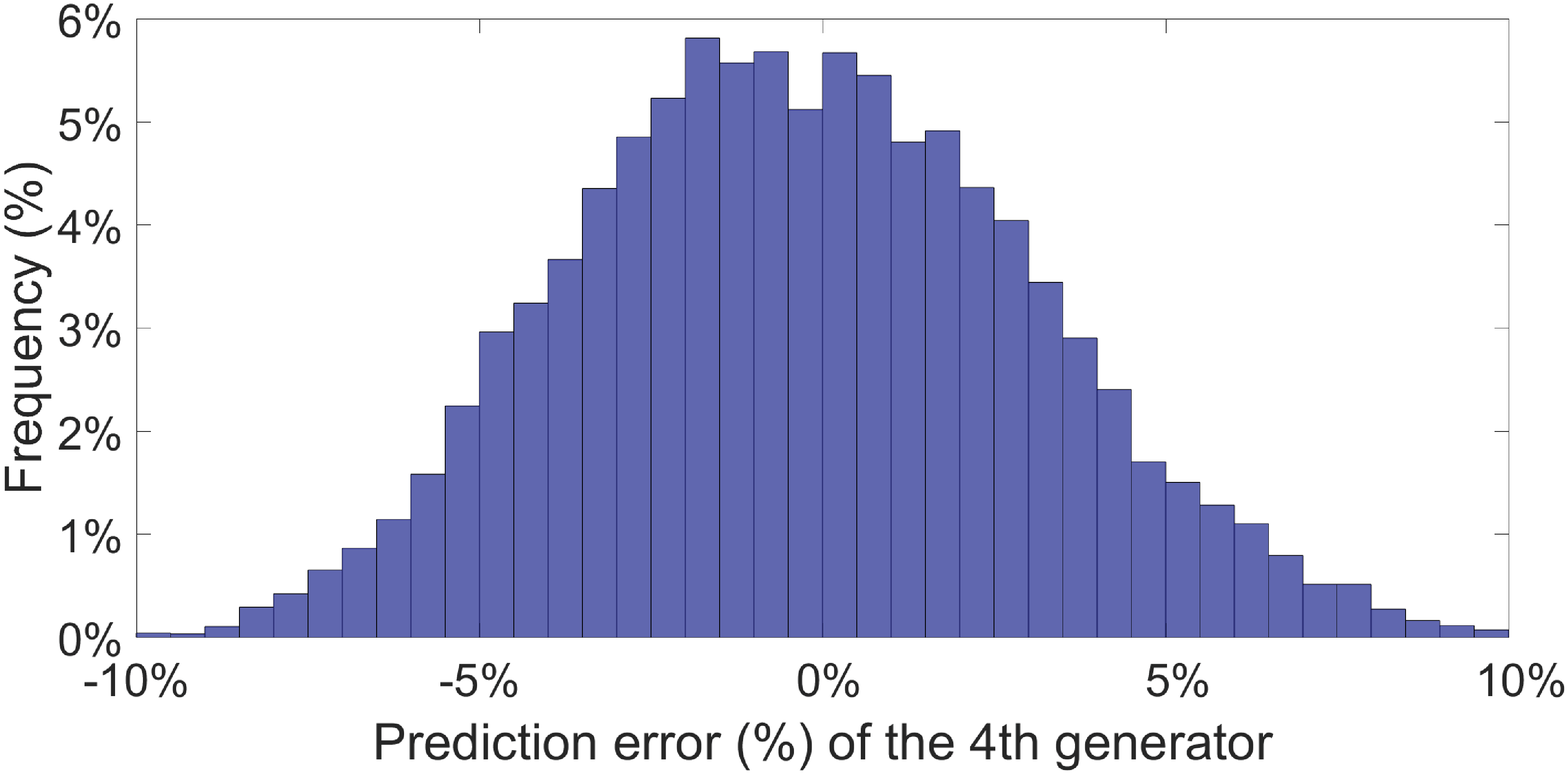}}
    \subfigure [] {\includegraphics[width = 0.3\textwidth]{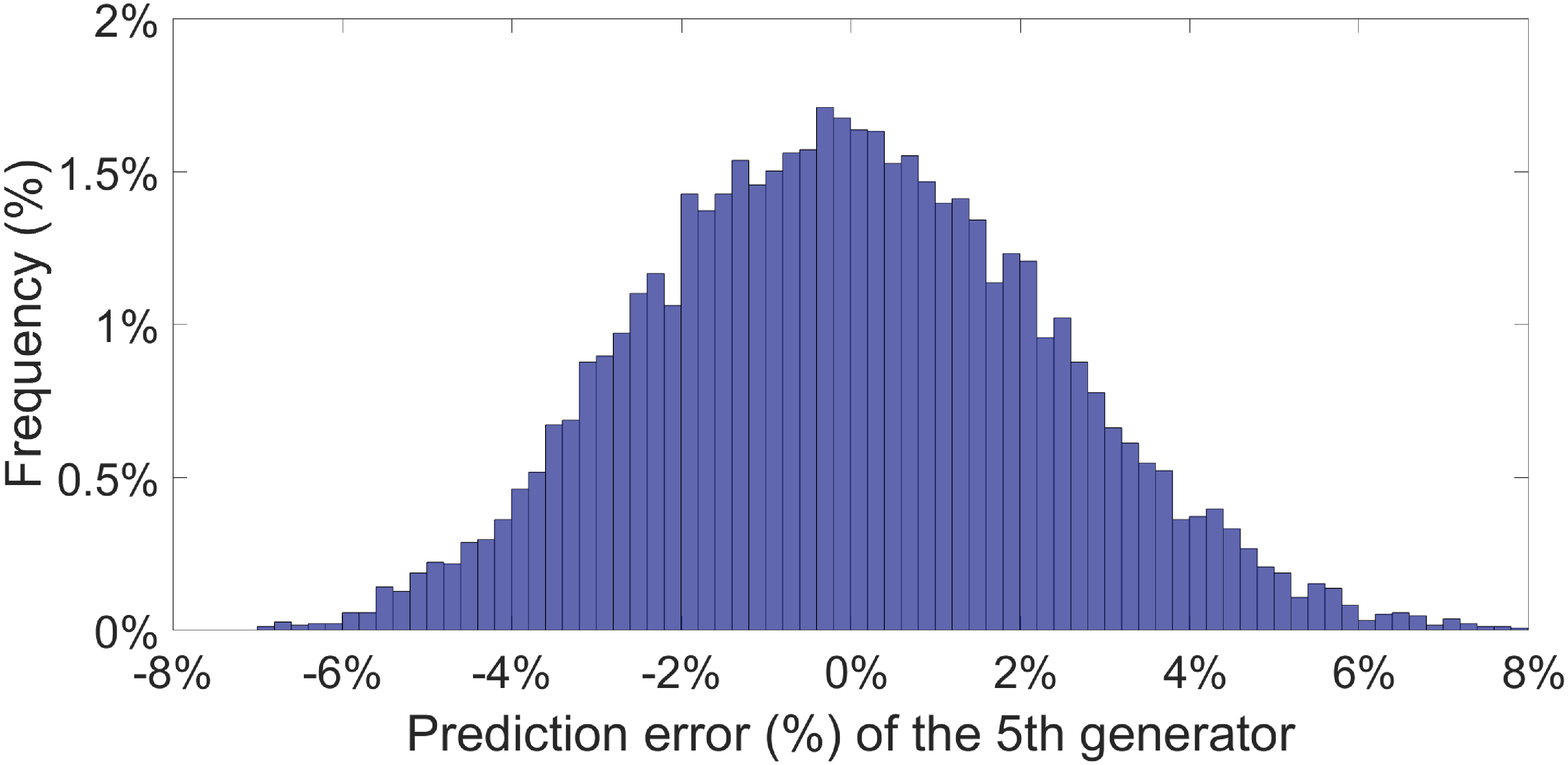}}
	\subfigure [] {\includegraphics[width = 0.3\textwidth]{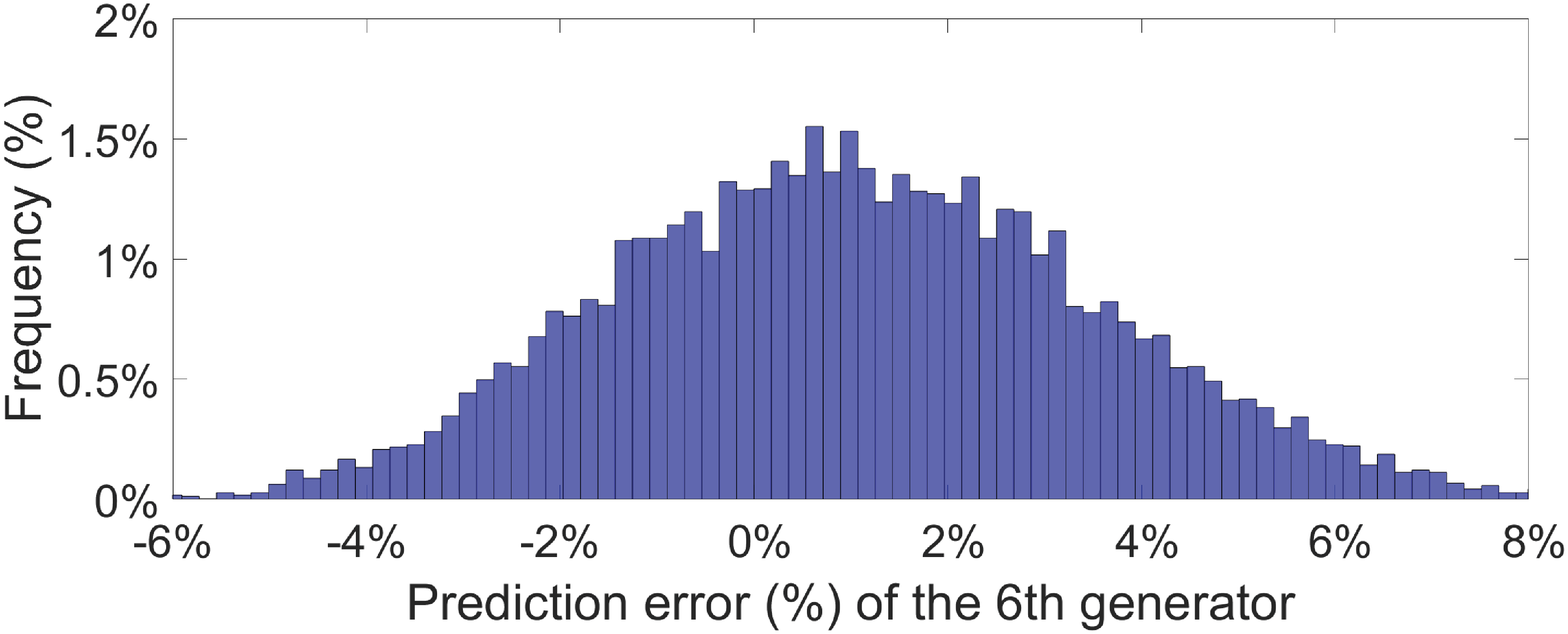}}	
	\caption{(a)-(f) represent the prediction errors of difference generators in the test stage under 10\% sample range of the 1st to the 6th generators.}
	\label{fig5}
\end{figure*}

\section{Proof of Lemma~\ref{lemma1}} \label{apx:proof_of_lemma_1}

For simplicity, let us use $y_{e}$ to denote the admittance of the branch $e\in E$ in the network, where $E$ is the set of all the branches. Since in the DC network, the resistance for each branch is negligible compared to the reactance and can therefore be set to 0 \cite{867153}, and consequently we have $y_e>0$, $\forall{e\in E}$.
It is easy to verify that the bus admittance matrix $B$ is given by:
$$B=A^{T}YA,$$
where $Y=\rm{diag}$$\left(y_e, e\in E \right)$ and $A$ is the $N_{\mbox{bran}}\times N_{\mbox{bus}}$ incidence matrix of the graph whose entries are given by, for $i\in \{1,2,\cdots,N_{\mbox{bran}}\}$ and $j \in \{1,2,\cdots,N_{\mbox{bus}}\}$,
\[ a_{ij} =
  \begin{cases}
    1,       & \quad \text{if } v_j \text{ is the positive end of }e_i;\\
    -1,  & \quad \text{if } v_j \text{ is the negative end of }e_i;\\
  0, & \quad \text{if } v_j \text{ is not joined with }e_i,
  \end{cases}
\]
where $N_{\mbox{bran}}$ is the number of the branches in the graph, i.e., number of elements in set $E$, and $N_{\mbox{bus}}$ is the number of the busses. We use $e_i$ to denote the $ith$ branch and $v_j$ as the $jth$ bus.

We first present a standard result from Graph Theory on the incidence matrix $A$.
\begin{lemma}[\cite{diestel2018graph}]
For a connected graph with $N_{\mbox{bus}}$ busses, the rank of the incidence matrix \textbf{A} is $N_{\mbox{bus}}-1$.
\end{lemma}
We present a simple proof in the following for completeness.
\begin{proof}

Let $C_j$ be the $j$th column of the above incidence matrix. Since every row of the matrix only has two no-zero entries i.e., +1 and -1, we have $\sum_{j=1}^{N_{\mbox{bus}}} C_j=0$, and consequently $rank(A) \leq N_{\mbox{bus}}-1$. Furthermore, for any linear combination of the column vectors of $A$, let $\sum_{j=1}^{N_{\mbox{bus}}} \alpha_i C_j=0$. Assume that the $kth$ column has a non-zero coefficient $\alpha_k$, then the column $C_k$ has non-zero entries in the rows which corresponds to the edges that starts or ends at the bus $v_k$. Since for each row there is only one $+1$ and one $-1$, suppose $a_{rk}\neq 0$, which means the bus $v_k$ is the starting or the ending bus of the edge $e_r$, we must have $a_{rk}$=$-a_{rl}$ for some $r$ as the graph is connected, which means $v_l$ is the other bus of the edge $e_r$. The linear relationship requires that $\alpha_k$=$\alpha_l$. Since the whole graph is connected, then we must have 
$$
    \alpha_i=\alpha_k, \forall j \in \{1,2 \dots N_{\mbox{bus}}\}.
$$

Hence the linear relationship is  $\alpha_k \left(\sum_{j=1}^{N_{\mbox{bus}}} C_j\right)=0$, which implies that $\sum_{j=1}^{N_{\mbox{bus}}} C_j=0$ is the only expression with a series not all zero $\{\alpha_1, \alpha_2, \dots, \alpha_{N_{\mbox{bus}}}\}$ and for any $N_{\mbox{bus}}-1$ combinations of column vectors there does not exist a series not all zero $\{\alpha_1, \alpha_2, \dots, \alpha_{N_{\mbox{bus}}-1}\}$ such that $\sum_{j=1}^{N_{\mbox{bus}}-1} \alpha_j C_j=0$. This complete the proof of  $rank(A)=N_{\mbox{bus}}-1$.
\end{proof}
To proceed with the proof of Lemma. \ref{lemma1}, we notice that
$$B=A^{T}YA=A^{T}\left(Y^{\frac{1}{2}}\right)^TY^{\frac{1}{2}}A,$$
where $\left(Y^{\frac{1}{2}}\right)^T=Y^{\frac{1}{2}}=\rm{diag}$ $\left(y^{\frac{1}{2}}_e,  e\in E\right)$. Then we have $$rank\left(B\right)=rank\left(A^T\left(Y^{\frac{1}{2}}\right)^TY^{\frac{1}{2}}A\right)=rank\left(Y^{\frac{1}{2}}A\right).$$
The above equality comes from that if a matrix $P$ is over the real numbers, then $rank\left(P^TP\right)=rank\left(P\right)$. Furthermore, notice that $Y^{\frac{1}{2}}$ is a diagonal matrix with diagonal elements be positive. Hence we have $Y^{\frac{1}{2}}$ is a full rank matrix. Then we get
$$rank(B)=rank(A)=N_{\mbox{bus}}-1.$$

Recall that the $\left(N_{\mbox{bus}}-1\right)\times \left(N_{\mbox{bus}}-1\right)$ matrix $\tilde{B}$ comes from deleting the row and column corresponding to the slack bus. Without loss of generality, let us assume the first row and column are removed from the original matrix. Let us use  $C^B_j$ to denote the $jth$ column of the matrix $B$.Let us use  $C^B_j$ to denote the $jth$ column of the matrix $B$.

 Let $\hat{B}$ be the matrix in which the first column is removed. Since we have $$C^B_1=-\sum^{N_{\mbox{bus}}-1}_{j=1}C^B_j,$$
removing this column dose not change the maximal number of linearly independent columns of $B$. Hence $rank(\hat{B})=rank(B)=N_{\mbox{bus}}-1$. Let us use $R^{\hat{B}}_i$ to denote the $ith$ row of matrix $\hat{B}$. Similarly we have
$$R^{\hat{B}}_1=-\sum^{N_{\mbox{bus}}-1}_{i=1}R^{\hat{B}}_i,$$
then removing this row does not change the maximal number of linear independent rows of matrix $\hat{B}$.
We conclude 
\begin{equation*}
rank\left(\tilde{B}\right)=rank\left(\hat{B}\right)=rank(B)=N_{\mbox{bus}}-1.  
\end{equation*} 
Consequently, the $(N_{\mbox{bus}}-1) \times (N_{\mbox{bus}}-1)$ matrix $\tilde{B}$ is a full rank matrix. 
This completes the proof of Lemma~\ref{lemma1}.

\section{The histogram of prediction errors of different generators for IEEE 30-bus case } \label{apx:histogram2}

As seen in Figure \ref{fig5}, the prediction errors of the generators labeled (b) to (f) basically fall on the range of $-5\%$ to $+5\%$, which indicates that our DNN model can find the relationship between the load and the proportion of electricity generated by each generator quite accurately. Notice that the first generator has a relative large prediction error. This comes from that the first one is used to settle the imbalance between the load and the existing power generation from the left $5$ generators, i.e., the bus where the first generator locates is the slack bus. Consequently, this generator will have a large distribution variance compared with others. The total cost is connected with the aggregated impacts of all generators with either positive or negative prediction errors,  and our DNN model provides a good cost saving performance with only about $0.17\%$ difference from optimum.

\begin{figure*}[h!]
    \centering
	\subfigure [IEEE 30-bus system] {\includegraphics[width = 0.45\textwidth]{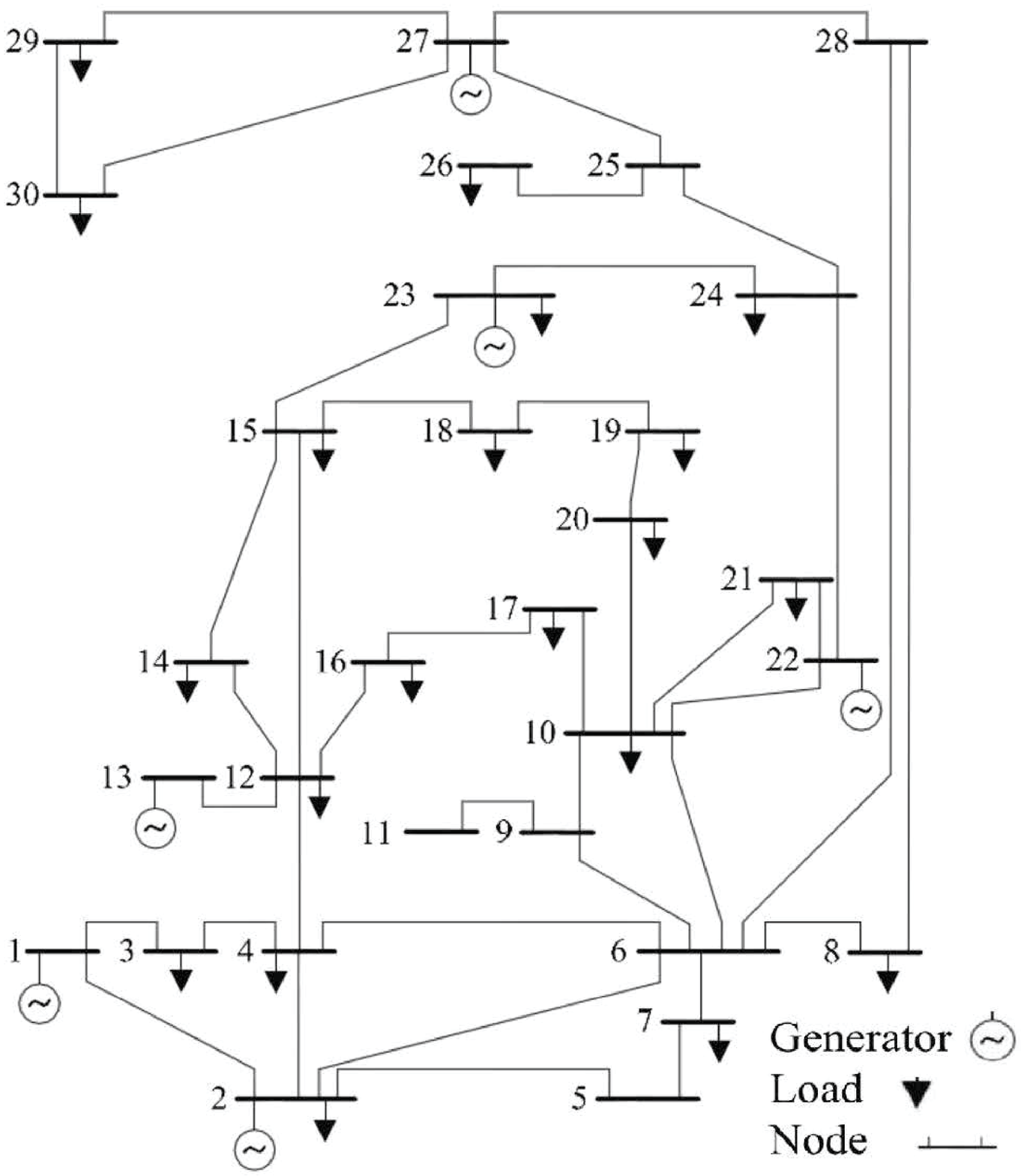}}
	\hspace{0.3in}
	\subfigure [IEEE 57-bus system] {\includegraphics[width = 0.45\textwidth]{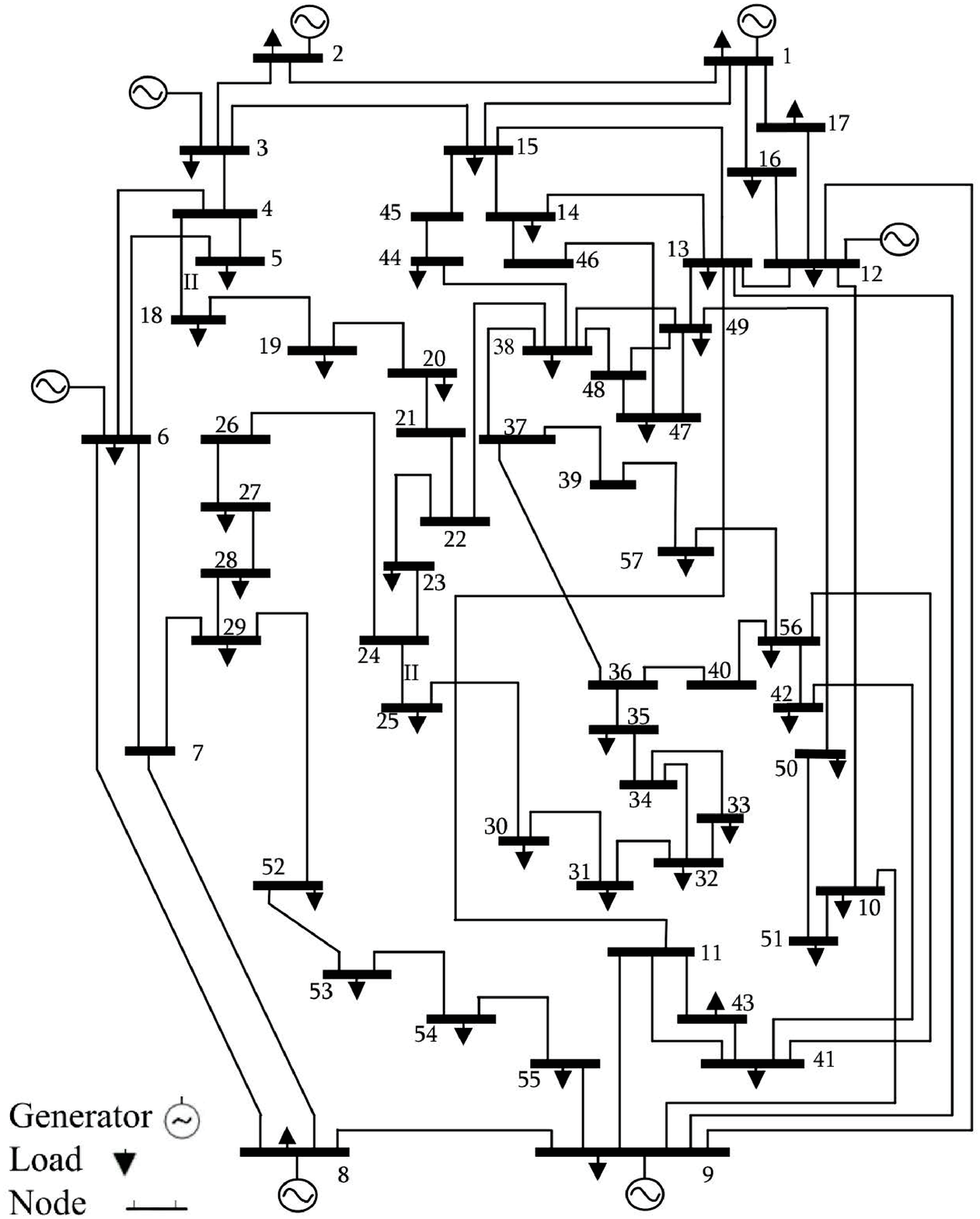}}
	\caption{(a)-(b) represent the illustration of IEEE 30-bus and IEEE 57-bus topology.}
	\label{fig6}
\end{figure*}

\section{The illustration of topology for the IEEE 30-bus and 57-bus system.} \label{apx:topology}
As shown in Fig. \ref{fig6}, the IEEE 30-bus test case consists of 30 buses, 6 generators, 41 branches and 20 loads. The IEEE 57-bus test case consists of 57 buses, 7 generators, 80 branches and 42 loads. The illustrations for the IEEE 118-bus and IEEE 300-bus cases can be found in \cite{4077121} and \cite{tpcwTrey6}, respectively.

\end{document}